\documentclass[a4paper,twoside]{article}

\usepackage{epsfig}
\usepackage{subcaption}
\usepackage{calc}
\usepackage{amssymb}
\usepackage{amstext}
\usepackage{amsmath}
\usepackage{amsthm}
\usepackage{multicol}
\usepackage{multirow}
\usepackage{pslatex}
\usepackage{apalike}
\usepackage[ruled,vlined,linesnumbered]{algorithm2e}
\usepackage[bottom]{footmisc}

\usepackage{url}
\usepackage[export]{adjustbox}
\usepackage{hyperref}

\usepackage{comment}
\usepackage{tabularx}
\usepackage{graphicx}
\usepackage{array}
\usepackage{xcolor}

\usepackage{enumitem}

\usepackage{listings}
\usepackage{color}
\usepackage{SCITEPRESS}     

\begin{document}

\title{Text-to-SQL based on Large Language Models and Database Keyword Search}

\author{
\authorname{Eduardo R. Nascimento\sup{1},
Caio Viktor S. Avila\sup{1,4},
Yenier T. Izquierdo\sup{1},
Grettel M. García\sup{1},
Lucas Feijó L. Andrade\sup{1},
Michelle S.P. Facina\sup{2},
Melissa Lemos\sup{1},
Marco A. Casanova\sup{1,3}} 
\affiliation{\sup{1}
Instituto Tecgraf, PUC-Rio, Rio de Janeiro, RJ, Brazil CEP 22451-900} 
\affiliation{\sup{2}
Petrobras, Rio de Janeiro, RJ, Brazil CEP 20231-030}
\affiliation{\sup{3}
Departamento de Informática, PUC-Rio, Rio de Janeiro, RJ, Brazil 
CEP 22451-900}
\affiliation{\sup{4}
Departamento de Computação, UFC, Fortaleza, Brazil,
CEP 60440-900}
\email{
\{rogerrsn,ytorres,ggarcia,lucasfeijo,melissa\}@tecgraf.puc-rio.br, caioviktor@alu.ufc.br,
michelle@petrobras.com.br,
casanova@inf.puc-rio.br
}
}


\abstract{Text-to-SQL prompt strategies 
based on Large Language Models (LLMs) 
achieve remarkable performance on well-known benchmarks.
However, when applied to real-world databases, 
their performance is significantly less than for these benchmarks,
especially for Natural Language (NL) questions 
requiring complex filters and joins to be processed.
This paper then proposes a strategy
to compile NL questions into SQL queries
that incorporates a dynamic few-shot examples strategy and 
leverages the services provided 
by a database keyword search (KwS) platform.
The paper details how the precision and recall 
of the schema-linking process are improved 
with the help of the examples provided 
and the keyword-matching service that the KwS platform offers.
Then, it shows how the KwS platform can be used 
to synthesize a view that captures the joins required
to process an input NL question
and thereby simplify the SQL query compilation step.
The paper includes experiments with a real-world relational database 
to assess the performance of the proposed strategy.
The experiments suggest that the strategy achieves
an accuracy on the real-world relational database 
that surpasses state-of-the-art approaches.
The paper concludes by discussing the results obtained.}

\onecolumn \maketitle \normalsize \setcounter{footnote}{0} \vfill

\section{\uppercase{Introduction}}
\label{section:introduction}

The \textit{Text-to-SQL} task is defined as
``\textit{given a relational database $D$ 
and a natural language (NL) sentence $Q_N$ 
that describes a question on $D$,
generate an SQL query $Q_{SQL}$ over $D$ that expresses $Q_N$}''
\cite{Katsogiannis2023,Kim2020}.

Numerous tools have addressed this task 
with relative success
\cite{Affolter2019,Katsogiannis2023,Kim2020,shi2024}
over well-known benchmarks,
such as Spider -- Yale Semantic Parsing and 
Text-to-SQL Challenge \cite{yu2018}
and BIRD -- BIg Bench for LaRge-scale Database 
Grounded Text-to-SQL Evaluation
\cite{li2023llm}.
The leaderboards of these benchmarks point to a firm trend: 
the best text-to-SQL tools 
are all based on Large Language Models (LLMs)
\cite{shi2024}.

Text-to-SQL tools must face several challenges.
To begin with, they must be able to process
NL questions that require multiple SQL constructs \cite{yu2018}.
For example, processing the NL question:
\begin{center}
\textit{``Which has more open orders, P-X or P-Y?''}
\end{center}
requires:
\begin{itemize}
    \item Recognizing that \textit{P-X} and \textit{ P-Y} are industrial installations;
    \item Joining installations and orders;
    \item Understanding what is an open order;
    \item Computing the number of open orders for each of the installations.
    \item Returning the installation with the largest number of open orders.
\end{itemize}

Omitting the details, the following SQL query
would answer the above NL question:

\begin{lstlisting}[language=SQL,
%    frame=lines,
%    frame=ltbr,
    keywordstyle=\color{black},
    numbersep=4pt,
    breaklines=true,
    showstringspaces=false,
    basicstyle={\scriptsize\ttfamily},
    upquote=true,
    emphstyle={\color{black}}]
    SELECT t.name, 
           COUNT(*) AS number_open_orders
     FROM Installation t JOIN Order o 
          ON t.code = o.installation_code
    WHERE (t.name = 'P-X' OR t.name = 'P-Y') 
      AND LOWER(o.status) LIKE LOWER ('%Open%')
    GROUP BY t.code
    ORDER BY number_open_orders DESC
    FETCH 1
\end{lstlisting}

This is an example of a challenging
NL question that the strategy proposed in this paper can compile into a correct SQL query.

In addition, \textit{real-world databases} raise a different set of challenges
for several reasons, among which:
\begin{enumerate}
\item The relational schema is often large,
in the number of tables, columns per table, and foreign keys --
which may lead to queries with many joins,
which are difficult to synthesize.
\item The relational schema 
is often an inappropriate specification
of the database from the point of view of the LLM --
the table and column names are often different from the terms
the users adopt to formulate their NL questions.
\item The data semantics are often complex; 
for example, some data values may encode enumerated domains,
which implies that the terms the users adopt to formulate their NL questions
must be mapped to this internal semantics.
\item Metadata and data are often ambiguous,
which influences the behavior of an LLM-based text-to-SQL tool,
leading to unexpected results.
\end{enumerate}
Indeed, the performance of some of the best LLM-based text-to-SQL tools
on real-world databases is significantly less than that observed 
for the Spider and BIRD benchmarks
\cite{nascimento2024a,Lei2024}.

This paper then addresses 
the \textit{real-world text-to-SQL problem},
which is the version of the text-to-SQL problem 
for real-world databases.
Albeit the original problem has been investigated for some time,
this version is considered far from solved,
as argued in \cite{Floratou2024,Lei2024}.

The first contribution of the paper is a novel strategy
to compile NL questions into SQL queries
that leverages the services provided by 
a database keyword search (KwS) platform,
called DANKE \cite{DANKE2021,Izquierdo2024}.
The proposed strategy is the first one to explore
a symbiotic combination of a KwS platform
and a prompt strategy to process NL questions.

Briefly, Section \ref{sec:schema-linking} details 
how the combination of DANKE's data dictionary  
with a dynamic few-shot examples strategy
improves the precision and recall of the schema-linking process, 
that is, the process of finding a set of tables 
that suffice to compile an input NL question.
Then, Section \ref{sec:query-compilation} shows 
how the SQL query compilation step is also improved by
calling DANKE to synthesize a view $V$ that captures 
the required joins to answer the input NL question $Q_N$,
and then calling an LLM to compile $Q_N$ into 
an SQL query $Q_{SQL}$ over $V$,
which can be remapped to the database schema 
with the help of the definition of $V$.

The second contribution of the paper 
is a set of experiments with a real-world benchmark 
to assess the performance of the proposed strategy.
The benchmark is built upon a relational database
with a challenging schema, which is in production at an energy company,
and a set of 100 NL questions carefully defined
to reflect the NL questions users submit
and to cover a wide range of SQL constructs
(Spider and BIRD, two of the familiar text-to-SQL benchmarks,
were not adopted for the reasons explained 
in Section \ref{sec:relatedbenchmarks}).
These new results, 
combined with results from \cite{nascimento2024a},
indicate that the proposed strategy performs significantly better
on the real-world benchmark than LangChain SQLQueryChain,
SQLCoder\footnote{\url{https://huggingface.co/defog/sqlcoder-34b-alpha}}, 
``C3 + ChatGPT + Zero-Shot'' \cite{dong2023c3}, and
``DIN-SQL + GPT-4'' \cite{pourreza2023dinsql}.

This paper is an extended version of \cite{nascimento2025}.

The paper is organized as follows.
Section \ref{sec:relatedwork} 
covers related work.
Section \ref{sec:KwS} 
describes the database keyword search platform 
adopted in the paper.
Section \ref{sec:prompt-strategy} 
details the proposed text-to-SQL strategy.
Section \ref{sec:experiments} 
presents the experiments,
including the real-world benchmark used.
Finally, Section \ref{sec:conclusions} 
contains the conclusions.

\section{\uppercase{Related Work}}
\label{sec:relatedwork}
\subsection{Text-to-SQL Datasets}
\label{sec:relatedbenchmarks}

The Spider -- Yale Semantic Parsing and Text-to-SQL Challenge \cite{yu2018}
defines 200 datasets,
covering 138 different domains,
for training and testing text-to-SQL tools.

For each database, 
Spider lists 20--50 hand-written NL questions 
and their SQL translations.
An NL question $S$, with an SQL translation $Q_N$,
is classified as easy, medium, hard, and extra-hard,
where the difficulty is based on 
the number of SQL constructs of $Q_N$ -- 
\texttt{GROUP BY}, \texttt{ORDER BY}, \texttt{INTERSECT}, 
nested sub-queries, column selections, 
and aggregators --
so that an NL query whose translation $Q_N$ 
contains more SQL constructs is
considered more complex.
The set of NL questions introduced in Section \ref{sec:questions}
follows this classification,
but does not consider extra-hard NL questions.

Spider proposes three evaluation metrics:
\textit{component matching} checks 
whether the components of the prediction and the ground-truth SQL queries match exactly; 
\textit{exact matching} measures 
whether the predicted SQL query as a whole is equivalent to 
the ground-truth SQL query;
\textit{execution accuracy} requires that
the predicted SQL query selects a list of gold values 
and fills them into the correct slots.
Section \ref{subsec:evaluation_metric} describes the metric
used in the experiments of this paper,
which is a variation of execution accuracy.

Most databases in Spider have very small schemas --
the largest five databases have between 16 and 25 tables, and 
about half have schemas with five tables or fewer.
Furthermore, all Spider NL questions 
are phrased in terms used in the database schemas.
These two limitations considerably reduce the difficulty of 
the text-to-SQL task.
Therefore, the results reported in the Spider leaderboard 
are biased toward databases with small schemas
and NL questions written in the schema vocabulary,
which is not what one finds in real-world databases.

Spider has two interesting variations. 
Spider-Syn \cite{gan2021a} is used to test how well text-to-SQL tools
handle synonym substitution,
and Spider-DK \cite{gan2021b} addressed testing how well 
text-to-SQL tools deal with domain knowledge.

BIRD -- BIg Bench for LaRge-scale Database Grounded Text-to-SQL Evaluation
\cite{li2023llm}
is a large-scale, cross-domain text-to-SQL benchmark in English. 
The dataset contains 12,751 text-to-SQL data pairs and 95 databases with a total size of 33.4 GB across 37 domains. 
However, BIRD still does not have many databases with large schemas --
of the 73 databases in the training dataset, 
only two have more than 25 tables,
and, of the 11 databases used for development, 
the largest one has only 13 tables. 
Again, all NL questions are phrased
in the terms used in the database schemas.

Finally, the \texttt{sql-create-context}\footnote{\url{https://huggingface.co/datasets/b-mc2/sql-create-context}} dataset 
also addresses the text-to-SQL task,
and was built from WikiSQL and Spider.
It contains 78,577 examples of NL questions, 
SQL \texttt{CREATE TABLE} statements, 
and SQL queries answering the questions. 
The \texttt{CREATE TABLE} statement provides context for the LLMs, 
without having to provide actual rows of data. 

Despite the availability of these benchmark datasets 
for the text-to-SQL task, and inspired by them,
Section \ref{sec:benchmark} describes a benchmark dataset
constructed specifically
to test strategies designed for the real-world text-to-SQL task.
The benchmark dataset consists of a relational database,
three sets of LLM-friendly views,
specified as described in Section \ref{sec:views},
and a set of 100 test NL questions
and their ground-truth SQL translations.
The database schema is inspired by a real-world schema and
is far more challenging than most of the database schemas
available in Spider or BIRD.
The database is populated with real data 
with a semantics which is sometimes not easily mapped
to the semantics of the terms the users adopt
(such as ``criticity\_level = 5'' encodes ``critical orders''),
which is a challenge for the text-to-SQL task
not captured by unpopulated databases, as in Spider.
Finally, the NL questions mimic those posed by real users,
and cover a wide range of SQL constructs
(see Table \ref{fig:datasets-stats}
in Section \ref{sec:questions}).
\subsection{Text-to-SQL Tools}
\label{sec:relatedtools}

A comprehensive survey of text-to-SQL strategies
can be found in \cite{shi2024},
including a discussion of benchmark datasets, 
prompt engineering, and fine-tuning methods,
partly covered in what follows.

The Spider Web site\footnote{\url{https://yale-lily.github.io/spider}} 
publishes a leaderboard with the best-performing text-to-SQL tools.
At the time of this writing, 
the top 5 tools achieved an accuracy that 
ranged from an impressive 85.3\% to 91.2\%
(two of the tools are not openly documented).
Four tools use GPT-4, as their names imply.
The three tools that provide detailed documentation
have an elaborate first prompt
that tries to select the tables and columns 
that best match the NL question.
Therefore, this first prompt is prone to failure
if the database schema induces a vocabulary
disconnected from the NL question terms.
This failure cannot be fixed by even more elaborate
prompts that try to match the schema and the NL question vocabularies,
but it should be addressed as proposed in this paper.

The BIRD Web site\footnote{\url{https://bird-bench.github.io}} 
also publishes a leaderboard with the best-performing tools.
At the time of this writing, 
out of the top 5 tools,
two use GPT-4, one uses CodeS-15B, one CodeS-7B, 
and one is not documented.
The sixth and seventh tools also use GPT-4, 
appear in the Spider leaderboard, 
and are well-documented.

The Awesome Text2SQL Web site\footnote{\url{https://github.com/eosphoros-ai/Awesome-Text2SQL}} 
lists the best-performing text-to-SQL tools 
on WikiSQL,
Spider (Exact Match and Exact Execution) and
BIRD (Valid Efficiency Score and Execution Accuracy).

The DB-GPT-Hub\footnote{\url{https://github.com/eosphoros-ai/DB-GPT-Hub}} 
is a project exploring how to use LLMs for text-to-SQL. 
It contains data collection, data preprocessing, model selection and building, and fine-tuning weights, including LLaMA-2,
and evaluating several LLMs fine-tuned for text-to-SQL.

Several text-to-SQL tools
were tested in \cite{nascimento2024a}
against the benchmark used in this paper --
SQLCoder, LangChain SQLQueryChain, C3, and DIN+SQL.

SQLCoder\footnote{\url{https://huggingface.co/defog/sqlcoder-34b-alpha}}
is a specialized text-to-SQL model,
open-sourced under the Apache-2 license.
The sqlcoder-34b-alpha model features 34B parameters and
was fine-tuned on a base CodeLlama model,
on more than 20,000 human-curated questions, 
classified as in Spider,
based on ten different schemas.

LangChain\footnote{\url{https://python.langchain.com}} 
is a generic framework that offers several pre-defined strategies 
to build and run SQL queries based on NL prompts.

``C3 + ChatGPT + Zero-Shot'' \cite{dong2023c3} (or briefly C3) is a prompt-based strategy,
originally defined for ChatGPT,
that uses only approximately 1,000 tokens per query and 
achieves a better performance than fine-tuning-based methods.
C3 has three key components:
\textit{Clear Prompting} (CP);
\textit{Calibration with Hints} (CH);
\textit{Consistent Output} (CO).
At the time of writing, C3 was the sixth strategy listed 
in the Spider leaderboard,
achieving 82.3\% in terms of execution accuracy on the test set.
It outperformed state-of-the-art fine-tuning-based approaches in execution accuracy on the test set.

``DIN-SQL + GPT-4'' \cite{pourreza2023dinsql} (or briefly DIN) 
uses only prompting techniques
and decomposes the text-to-SQL task into four steps:
\textit{schema linking}; 
\textit{query classification and decomposition}; 
\textit{SQL generation}; and 
\textit{self-correction}.
When released, DIN was the top-performing tool 
listed in the Spider leaderboard,
achieving 85.3\% in terms of execution accuracy.

Despite the impressive results of 
C3 and DIN on Spider,
and of SQLCoder on a specific benchmark,
the performance of these tools
on the benchmark used in this paper 
was significantly lower \cite{nascimento2024a},
and much less than that of the strategy 
described in Section \ref{sec:prompt-strategy}.
A similar remark applies to LangChain SQLQueryChain,
whose results are shown in Line 1 of Table \ref{tab:results}.

\subsection{Retrieval-Augmented and Dynamic Few-shot Examples Prompting}
\label{sec:rag_related}

Retrieval-Augmented Generation (RAG), introduced in \cite{Lewis2020},
is a strategy to incorporate data from external sources.
This process ensures that the responses are grounded 
in retrieved evidence, 
thereby significantly enhancing the accuracy and relevance of the output.
There is an extensive literature on RAG.
A recent survey \cite{gao2024}
classified RAG strategies into
\textit{naive}, \textit{advanced}, and \textit{modular} RAG.
Naive RAG follows the traditional process 
that includes indexing, retrieval, and generation of document ``chunks''.
Advanced RAG introduces various methods to optimize retrieval. 
Modular RAG integrates strategies 
to enhance functional modules, 
such as incorporating a search module 
for similarity retrieval and applying 
a fine-tuning approach in the retriever.

As for text-to-SQL, recent references 
include a RAG technique \cite{panda2024} 
to retrieve table and column descriptions from a metadata store 
that are related to the NL question,
based on similarity search.

LangChain offers a dynamic few-shot examples prompting technique\footnote{https://python.langchain.com/v0.1/docs/use\_cases/sql/ prompting/} also based on similarity search.
Given an NL question $Q_N$,
the prompting strategy includes the examples most relevant to $Q_N$,
retrieved by a similarity search between $Q_N$ and 
a set of examples previously stored in a vector database.

A retrieval-augmented prompting method for a LLM-based text-to-SQL framework
is proposed in \cite{guo2023}, involving sample-aware prompting 
and a dynamic revision chain. 
The method uses two strategies to retrieve questions 
sharing similar intents with input questions. 
Firstly, using LLMs, the method simplifies the original questions, 
unifying the syntax and thereby clarifying the users' intentions. 
To generate executable and accurate SQL queries without human intervention, 
the method incorporates a dynamic revision chain,
which iteratively adapts fine-grained feedback 
from the previously generated SQL queries.

A similar strategy is proposed in \cite{coelho2024},
that also describes a technique 
to create synthetic datasets with sets of examples
$(Q,S)$ where $Q_N$ is an NL question and $S$ is its SQL translation.

\section{\uppercase{A Database Keyword Query Processing Tool}}
\label{sec:KwS}

DANKE is the keyword search platform currently 
deployed for the industrial database described 
in Section \ref{sec:benchmark}
and used for the experiments.
The reader is referred to ~\cite{DANKE2021,Izquierdo2024}
for the details of the platform.

DANKE operates over both relational databases and RDF datasets, 
and is designed to compile a keyword query into 
an SQL or SPARQL query that returns the best data matches. 
For simplicity, the description that follows 
uses the relational terminology.

DANKE's architecture comprises three main components: 
(1) \textit{Storage Module}; 
(2) \textit{Preparation Module}; and 
(3) \textit{Data and Knowledge Extraction Module}. 

The \textit{Storage Module} houses 
a centralized relational database,
constructed from various data sources.
The database is described by a \textit{conceptual schema}, 
treated in what follows as a relational schema,
again for simplicity.

The \textit{Storage Module} also holds the data indices required 
to support the keyword search service.
The indexing process is enriched to create 
a \textit{keyword dictionary} containing:
\begin{itemize}
    \item for each table $T$,
    an entry of the form $(T,T_S)$, 
    where $T_S$ is a list of terms users adopt to refer to table $T$.
    \item for each column $A$ of a table $T$,
    an entry of the form $(A,T,A_S)$,
    where $A_S$ is a list of terms that users adopt to refer to 
    column $A$ in the context of table $T$.
    \item for each indexed value $v$,
    an entry of the form $(v,T,A)$,
    where $T[A]$ is the table/column where $v$ occurs.
\end{itemize}

The \textit{Preparation Module} has tools 
for creating the conceptual schema and 
for constructing and updating the centralized database 
through a pipeline typical of a data integration process. 
The conceptual schema is defined
by de-normalizing the relational schemas of the underlying databases 
and indicating which columns will have their values indexed.

The \textit{Data and Knowledge Extraction Module} 
has two main sub-modules, 
\textit{Query Compilation}
and \textit{Query Processing}.

Given a keyword query, 
represented by a list of keywords,
the \textit{Query Compilation Module} has three major steps:
\begin{enumerate}
\item \textit{(Matching Discovery)} 
Match each keyword in the keyword query 
with table and column names or data values 
in the keyword dictionary.
\item \textit{(Matching Optimization)} 
Select the most relevant matches.
\item \textit{(Conceptual Query Compilation)} 
Compile a \textit{conceptual query} over the conceptual schema
from the most relevant matches.
\end{enumerate}

The \textit{Query Processing Module}, in turn, has two major steps:
\begin{enumerate}
\item \textit{(Query Compilation)} 
Compile the conceptual query into an SQL query.
\item \textit{(Query Execution)}
Submit the SQL query for execution,
collect the results, and display them to the user.
\end{enumerate}

Let $R$ be the referential dependencies diagram
of the database schema in question,
where the nodes of $R$ are the tables
and there is an edge between nodes $t$ and $u$ 
iff there is a foreign key from $t$ to $u$ or vice-versa.
Given a set of keywords $K$,
let $T_K$ be a set of table schemes 
whose instances match the largest set of keywords in $K$. 
The conceptual query compilation step 
first constructs a Steiner tree $S_K$ of $R$ 
whose end nodes are the set $T_K$.
This is the central point since it guarantees that
the final SQL query will not return unconnected data,
as explored in detail in \cite{Garcia2017}.
If $R$ is connected, then it is always possible to construct
one such Steiner tree;
overwise, one would have to find a Steiner forest 
to cover all tables in $T_K$. 

Using the Steiner tree,
the Query Compilation step compiles the keyword query 
into an SQL query that includes restriction clauses 
representing the keyword matches and join clauses connecting 
the restriction clauses. 
Without such join clauses, an answer would be a disconnected 
set of tuples, 
which hardly makes sense. 
The generation of the join clauses 
uses the Steiner tree edges.

Lastly, DANKE's internal API was expanded
to support the text-to-SQL strategy described in Section \ref{sec:prompt-strategy}. 
Briefly, it now offers the following services:
\begin{itemize}
    \item \textit{Keyword Match Service:} receives a set $K$ of keywords and
    returns the set $K_M$ of pairs $(k,d_k)$
    such that $k \in K$ and $d_k$ is the dictionary entry that best matches $k$,
    using the matching optimization heuristic mentioned above.
    The dictionary entry $d_k$ will be called the \textit{data associated} with $k$.
    \item \textit{View Synthesis Service:} receives a set $S'$ of tables
    and returns a view $V$ that best joins all tables in $S'$,
    using the Steiner tree optimization heuristic mentioned above.
\end{itemize}

\section{\uppercase{A Strategy for the Text-to-SQL Task}}
\label{sec:prompt-strategy}

\subsection{Outline of the Proposed Strategy}
\label{sec:strategy-outline}

\begin{figure}[!ht]
    \centering
    \includegraphics[width=\linewidth]{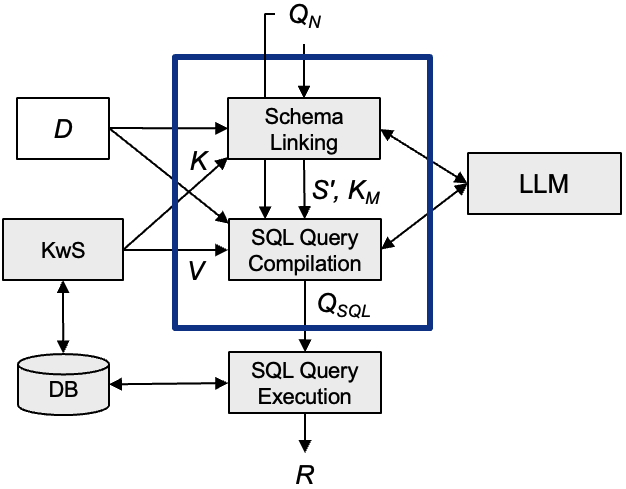}
    \caption{Proposed strategy.}
    \label{fig:prompt-strategy}
\end{figure}

Briefly, the proposed strategy comprises two modules, 
\textit{schema linking} and \textit{SQL query compilation}, 
as typical of text-to-SQL prompt strategies.
Figure \ref{fig:prompt-strategy} summarizes the proposed strategy,
leaving the details to the next sections.

The two modules run under LangChain.
They use a dynamic few-shot examples strategy 
that retrieves a set of samples from a \textit{synthetic dataset} $D$,
indexed with the help of the FAISS similarity search  library\footnote{https://ai.meta.com/tools/faiss/}.
The key point is the use of services provided by DANKE
to enhance schema linking and simplify SQL compilation,
as explained in the following sections.
In particular, DANKE will generate a single SQL view
containing all data and encapsulating all joins
necessary to answer the input NL question.

The current implementation runs in-house: 
LangChain, FAISS, DANKE, and Oracle.
The experiments used the OpenAI GPT-4
and its variations, as detailed in Section \ref{sec:experiments}.

\subsection{Synthetic Dataset Construction}
\label{sec:synthetic-dataset}

Let $DB$ be a relational database with schema $S$.
A \textit{synthetic dataset} $D$ for $DB$
contains pairs $(Q_N,Q_{SQL})$,
where $Q_N$ is an NL question and $Q_{SQL}$ is its SQL translation.
Such pairs should provide examples that help the LLM understand
how the database schema is structured,  
how the user's terms map to terms of the database schema,
and how NL language constructions map to data values.

The synthetic dataset construction process
repeatedly calls Algorithm \ref{alg:GenerateExample} 
to generate as many pairs $(Q_N,Q_{SQL})$ as desired.
The parameter $n$ is set in each call to determine 
how many tables the SQL query should involve.

\begin{algorithm}[!ht]
\SetAlgoLined
 \caption{Generating examples for the synthetic dataset.}
 \label{alg:GenerateExample} 
 \KwData{the number $n$ of tables to select, the database $DB$, the database schema $S$, and the database documentation $DB_{doc}$, if available.}
 \KwResult{a pair $(Q_N,Q_{SQL})$ 
    where $Q_N$ is an NL question and $Q_{SQL}$ 
    is the corresponding SQL query.}
\SetKwProg{Fn}{Function}{:}{}
\Fn{CreateExample($n,DB,S,DB_{doc}$)}{
    $T$ $\gets$ SelectTables($n,S$)\;
    $C$ $\gets$ 
    SelectColumns($T,S$)\;
    $L$ $\gets$ 
    CreateDDL($T,C$)\;
    $Q_{N'}$ $\gets$ CreateQuestion($L$)\;
    $Q_{SQL}$ $\gets$ 
    GenerateSQL($Q_{N'},L$)\;
    $Q_N$ $\gets$ ImproveQuestion($Q_{N'},S,DB_{doc}$)\;
    \KwRet{$(Q_N,Q_{SQL})$}\;
}
\end{algorithm}

Step 1 (on Line 2) selects a set $T$ of $n$ tables 
from the database schema $S$. 
The selection process employs a weighted random distribution, 
which reflects the likelihood of 
each table being chosen by an average user.
Note that users may choose some tables more often than others,
which justifies employing a weighted random distribution,
obtained from the users' access log.
    
Step 2 (on Line 3) 
selects column pairs for each table chosen in Step 1. 
The first column selected is always the primary key of the table, 
and the second column is chosen based on the weighted random 
distribution of each column in the database schema $S$.
    
Step 3 (on Line 4) creates a simplified Data Definition Language (DDL) 
statement $L$, encompassing only the columns and tables involved. 
Column and table names are renamed to their respective names 
in the conceptual schema views (see Section \ref{sec:views}).
    
Step 4 (on Line 5) creates an NL question $Q_{N'}$ 
by prompting GPT-4 with the simplified DDL statement $L$ 
and sample values of each column from the database $DB$. 
In addition, the prompt includes the type of restriction 
to be incorporated into the NL question, 
which depends on the data type of each column. 
For example, numeric-type columns can be used 
to create queries with aggregations.   
Finally, the prompt includes instructions 
that indicate that $Q_{N'}$ must be generated 
in the database vocabulary; that is, 
the table and column names must be kept 
to facilitate the generation of the SQL corresponding to the NL question.
    
Step 5 (on Line 6) calls GPT-4 to translate $Q_{N'}$ 
into an SQL query that responds to the NL question 
by providing $Q_{N'}$ and $L$. 
This process is facilitated by the fact that 
the SQL query should use the database schema vocabulary; 
that is, the prompt provides clues about 
which tables, columns, and values are involved.

The key point is to explore the column type
to decide which SQL construct must be used to express
a restriction for the column.
For example,
given a string $B$,
if column \texttt{INSTALLATION\_NAME} were of type \texttt{STRING}
and not a key,
the prompt would guide GPT-4 to create
a restriction of the form 
\begin{center}
\texttt{INSTALLATION\_NAME LIKE `\%$B$\%'} 
\end{center}
However, if column \texttt{INSTALLATION\_NAME} were a key,
then the prompt would guide GPT-4 to create
a restriction of the form 
\begin{center}
\texttt{INSTALLATION\_NAME = `$B$'}
\end{center}

Finally, Step 6 (on Line 7) calls GPT-4 
to translate $Q_{N'}$ into an improved NL question $Q_N$, 
using the database documentation $DB_{doc}$, 
which includes the description of each column and table, 
along with synonyms. 
During this step, the LLM is instructed 
to rephrase the NL question by translating from 
the database schema vocabulary to the user's vocabulary, 
preserving the original NL question intent.

\subsection{Schema Linking}
\label{sec:schema-linking}

Let $DB$ be a relational database with schema $S$
and $D$ be the synthetic dataset created for $DB$.
Let $Q_N$ be an NL question over $S$.

\begin{figure}[!ht]
    \centering
    \includegraphics[width=\linewidth]{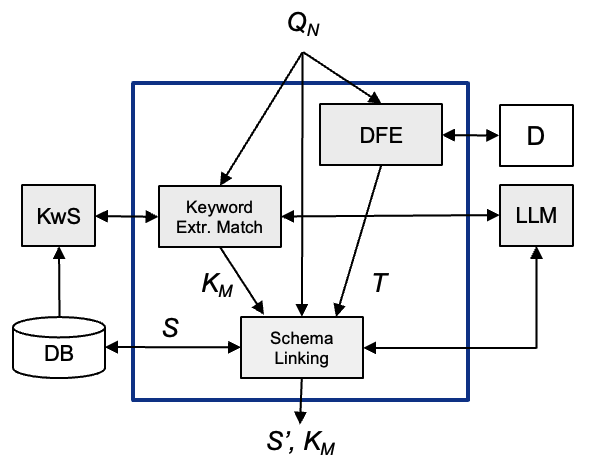}
    \caption{Schema linking module.}
    \label{fig:schema_linking}
\end{figure}

The schema linking module 
primarily finds a minimal set $S' \subset S$ 
such that $S'$ has all tables in $S$ required to answer $Q_N$.
It has the following major components
(see Figure \ref{fig:schema_linking}):

\vspace{1mm}
\begin{description}
    \item \textit{Keyword Extraction and Matching}
    \begin{enumerate}
    \item Receives as input an NL question $Q_N$.
    \item Calls the LLM to extract a set $K$ of keywords from $Q_N$.
    \item Calls the DANKE Keyword Matching service to match $K$ with the dictionary,
    creating a final set $K_M$ of keywords and associated data.
    \item Returns $K_M$.
    \end{enumerate}
    \item \textit{Dynamic Few-shot Examples Retrieval (DFE)} 
    \begin{enumerate}
    \item Receives as input an NL question $Q_N$.
    \item Retrieves from the synthetic dataset $D$
    a set of $k$ examples whose NL questions 
    are most similar to $Q_N$, 
    generating  
    \begin{center}
    $L = [(Q_1,SQL_1),...,(Q_k,SQL_k)]$
    \end{center}
    \item Creates a list $T$ of pairs by 
    retaining only the table names 
    in the \texttt{FROM} clauses, that is,
    \begin{center}
    $T = [(Q_1,F_1),...,(Q_k,F_k)]$
    \end{center}
    where $F_i$ is the set of tables in the 
    \texttt{FROM} clause of $SQL_i$.
    \item Returns $T$.
    \end{enumerate}
    \item \textit{Schema Linking}
    \begin{enumerate}
    \item Receives as input an NL question $Q_N$, 
    a set $K_M$ of keywords and associated data, 
    and a list $T$ as above.
    \item Retrieves the set of tables in $S$ and their columns.
    \item Calls the LLM to create $S'$ 
    prompted by $Q_N$, $K_M$, $S$, and $T$.
    \item Returns $S'$ and $K_M$.
    \end{enumerate}
\end{description}
\vspace{-2mm}

\subsection{SQL Query Compilation}
\label{sec:query-compilation}

\begin{figure}[!ht]
    \centering
    \includegraphics[width=\linewidth]{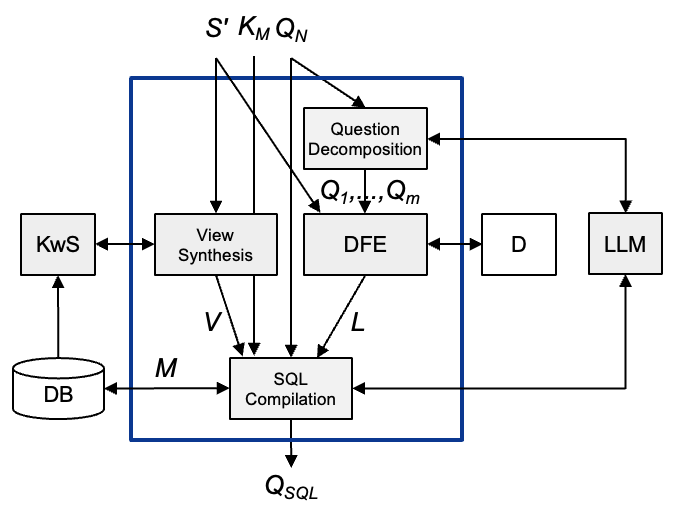}
    \caption{SQL query compilation module.}
    \label{fig:sql-compilation}
\end{figure}

The SQL query compilation module
receives as input the NL question $Q_N$, 
the set of tables $S'$, and 
the set $K_M$ of keywords and associated data,
and returns an SQL query $Q_{SQL}$.
It has the following major components
(see Figure \ref{fig:sql-compilation}):
\begin{description}
    \item \textit{View Synthesis} 
    \begin{enumerate}
    \item Receives as input a set of tables $S'$.
    \item Calls the DANKE View Synthesis service to synthesize a view $V$ 
    that joins the tables in $S'$.
    \item Returns $V$.
    \end{enumerate}
    \item \textit{Question Decomposition} 
    \begin{enumerate}
    \item Receives as input an NL question $Q_N$.
    \item Decomposes $Q_N$ into sub-questions $Q_1,...,Q_m$.
    \item Returns $Q_1,...,Q_m$.
    \end{enumerate}
    \item \textit{Dynamic Few-shot Examples Retrieval (DFE)} 
    \begin{enumerate}
    \item Receives as input a list of NL questions $Q_1,...,Q_m$.
    \item Let $p = \lceil k/m \rceil$.
    For each $i \in [1,m]$, 
    retrieves from the synthetic dataset $D$
    a set of $p$ examples whose NL questions 
    are most similar to $Q_i$ and whose SQL queries are over $S'$, 
    generating a list  
    \begin{center}
    $L_i = [(Q_{i_1},SQL_{i_1}),...,(Q_{i_p},SQL_{i_p})]$
    \end{center}
    in decreasing order of similarity of $Q_{i_j}$ to $Q_i$.
    \item Creates the final list $L$, with $k$ elements,
    by intercalating the lists $L_i$ and retaining the top-k pairs.
    \item Returns $L$.
    \end{enumerate}
\vspace{1mm}
    \item \textit{SQL Compilation}
    \begin{enumerate}
    \item Receives as input a view $V$,  
    a set $K_M$ of keywords and associated data, 
    an NL question $Q_N$,  
    and a list $L$ as above.
    \item In each SQL query $SQL_{i_j}$ in $L$, 
    replaces all tables in the FROM clauses by $V$, creating a new list $L'$.
    \item Retrieves from $DB$ a set $M$ of row samples of $V$.
    \item Calls the LLM to compile $Q_N$ into an SQL query $Q_{SQL}$ over $V$, 
    when prompted with $Q_N$, $V$, $K_M$, $M$ and $L'$.
    \item Returns $Q_{SQL}$.
    \end{enumerate}
\end{description}

\subsection{Limitations and Examples}
\label{sec:example}

If the proposed text-to-SQL strategy can compile 
an NL question $Q_N$ into an SQL query $Q_{SQL}$, 
then $Q_{SQL}$ is such that:
\begin{itemize}
\item $Q_{SQL}$ 
is defined over a single table $V$.
\item $V$ is a view defined by an SQL query
over a single table or a block of equijoin clauses, 
with no WHERE clause and no optional clauses
(GROUP BY, HAVING, ORDER BY, and LIMIT clauses).
\end{itemize}

These conditions
reflect the way the text-to-SQL strategy is structured.
Indeed, first, observe that, 
in the \textit{SQL Query Compilation} module, 
the \textit{View Synthesis} step calls 
DANKE's \textit{View Synthesis Service}
to create a view $V$,
defined by a set of equijoin clauses
over the set of tables $S'$ passed by the schema linking module.
This step improves the (manual) approach 
proposed in \cite{nascimento2024b}.

Then, the \textit{SQL Compilation} step prompts the LLM with view $V$ 
to generate the SQL query.  
View $V$ then facilitates 
the translation of an NL question into an SQL query
since the LLM no longer needs to discover 
which joins to include in the SQL query.

The predicted SQL queries and the views
in Tables \ref{fig:sample-medium-questions} and 
\ref{fig:sample-complex-questions} at the end of the paper
provide examples of such NL questions, SQL queries, and views,
two of which are discussed in more detail in what follows.

As a very simple example, 
consider Question 24 in Table \ref{fig:sample-medium-questions},
which requires joining tables \texttt{Recommendation} 
and \texttt{Installation}, 
as its ground-truth SQL query indicates.
The schema linking module can compute
that Question 24 requires these two tables.
Then, the \textit{View Synthesis} step 
calls DANKE, which receives the relational schema
(see Table \ref{fig:indb}) and these two tables, 
finds the required join,
and synthesizes the following view 
(some details of the view definition are omitted here for brevity):

\begin{lstlisting}[language=SQL,
%    frame=lines,
%    frame=ltbr,
    keywordstyle=\color{black},
    numbersep=4pt,
    breaklines=true,
    showstringspaces=false,
    basicstyle={\ttfamily},
    upquote=true,
    emphstyle={\color{black}}]
CREATE VIEW 
   Recommendation_Installation AS 
   SELECT r.id AS 
          Recommendation_id,
          r.situation AS 
          Recommendation_situation,
          ...
     FROM Recommendation r 
     JOIN Installation p
          ON r.installation_name =
             p.name
\end{lstlisting}

\noindent
As a result, 
the SQL compilation step
generates an SQL query without explicitly 
including this join
(see the predicted SQL query for Question 24 in 
Table \ref{fig:sample-medium-questions}).

The SQL compilation step 
prompts this view to the LLM as 
if it were a table in DDL format:

\begin{lstlisting}[language=SQL,
%    frame=lines,
%    frame=ltbr,
    keywordstyle=\color{black},
    numbersep=4pt,
    breaklines=true,
    showstringspaces=false,
    basicstyle={\ttfamily},
    upquote=true,
    emphstyle={\color{black}}]
CREATE TABLE 
   Recommendation_Installation 
   (Recommendation_id,
    Recommendation_situation,...)
\end{lstlisting}

However, this very simple example does not
fully illustrate the power of using DANKE
to join any number of tables.
A slightly more complex example goes as follows.

Consider Question 93 in Table \ref{fig:sample-complex-questions},
which requires joining tables \texttt{Maintenance\_request},
\texttt{Maintenance\_recommendation}, and 
\texttt{Maintenance\_order},
as its ground-truth SQL query indicates.
Again, the schema linking module can compute
that Question 93 requires these three tables.
Then, the \textit{View Synthesis} step 
calls DANKE, which receives the relational schema and these three tables, 
finds the required two joins,
and synthesizes the following view 
(again, some details of the view definition are omitted here for brevity):

\begin{lstlisting}[language=SQL,
%    frame=lines,
%    frame=ltbr,
    keywordstyle=\color{black},
    numbersep=4pt,
    breaklines=true,
    showstringspaces=false,
    basicstyle={\ttfamily},
    upquote=true,
    emphstyle={\color{black}}]
CREATE VIEW 
 Request_Recommendation_Order AS 
 SELECT m.id AS Request_id,
        r.id AS Recommendation_id,
        o.id AS Order_id,
            ...
 FROM Maintenance_request m 
 JOIN Maintenance_recommendation r
      ON m.id = r.note_id
 JOIN Maintenance_order o
      ON r.order_id = o.id
\end{lstlisting}

\noindent
As a result, the SQL compilation step
generates an SQL query without explicitly including these joins
(see the predicted SQL query for Question 93 in 
Table \ref{fig:sample-complex-questions}).

The SQL compilation step prompts this view to the LLM as 
if it were a table in DDL format:

\begin{lstlisting}[language=SQL,
%    frame=lines,
%    frame=ltbr,
    keywordstyle=\color{black},
    numbersep=4pt,
    breaklines=true,
    showstringspaces=false,
    basicstyle={\ttfamily},
    upquote=true,
    emphstyle={\color{black}}]
CREATE TABLE 
   Request_Recommendation_Order 
   (Request_id,
    Recommendation_id,
    Order_id,...)
\end{lstlisting}

\section{\uppercase{Experiments}}
\label{sec:experiments}

\subsection{A Benchmark Dataset}
\label{sec:benchmark}

This section describes a benchmark
to help investigate the real-world text-to-SQL task. 
The benchmark consists of a relational database,
a set of 100 test NL questions
and their SQL \textit{ground-truth} translations,
and a set of \textit{partially extended views}.

\subsubsection{The Relational Database}
\label{sec:relational-database}

The selected database is a real-world relational database (in Oracle) 
that stores data related to the integrity management 
of an energy company's industrial assets.
The relational schema of the adopted database contains 
27 relational tables with, in total, 
585 columns and 30 foreign keys (some multi-column), 
where the largest table has 81 columns.

\begin{figure*}[ht]
\centering
\includegraphics[width=0.90\linewidth]{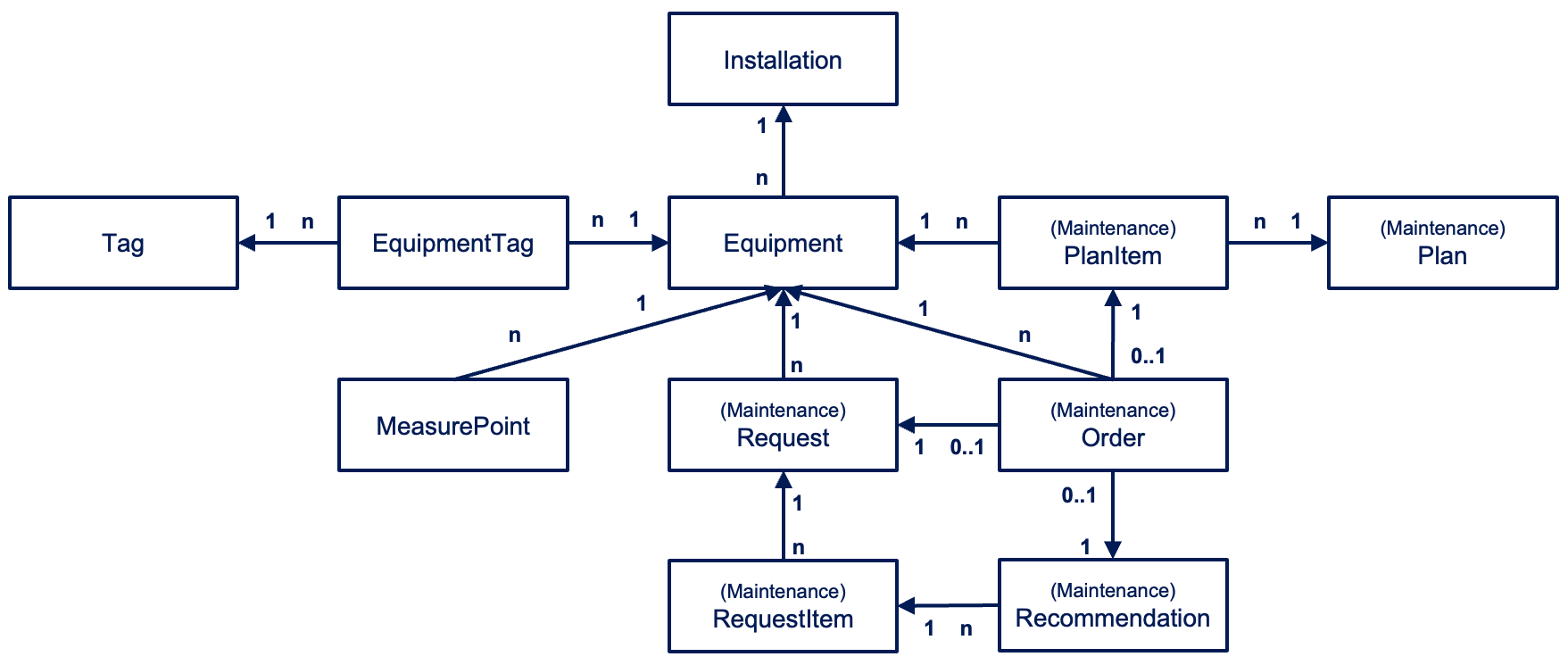}
\caption{The referential dependencies diagram of a simplified and anonymized version of the industrial database schema.}
\label{fig:indb}
\end{figure*}

Figure \ref{fig:indb}
shows the referential dependencies diagram of 
a much-simplified and anonymized version of the relational schema
of the real-world database,
where an arrow represents a foreign key
and points to the referenced table, as usual.
Note that the diagram is a connected graph,
which implies that,
given any set of tables $T$ of the relational schema,
it is always possible to create a Steiner tree of the diagram
that covers all tables in $T$.
This implies that Step 1 of the SQL Query compilation process
will always succeed in creating the required view.

\subsubsection{The Set of Test Questions and their Ground-Truth SQL Translations}
\label{sec:questions}

The benchmark contains a set of 100 NL questions
that consider the terms and questions experts use 
when requesting information related to 
the maintenance and integrity processes.
The ground-truth SQL queries
were manually defined over the conceptual schema views 
so that the execution of a ground-truth SQL query 
returns the expected answer to the corresponding NL question.

An NL question is classified into 
\textit{simple}, \textit{medium}, and \textit{complex},
based on the complexity
of its corresponding ground-truth SQL query,
as in the Spider benchmark
(extra-hard questions were not considered).
The set $L$ contains 33 simple, 33 medium, 
and 34 complex NL questions, 
with the basic statistics shown 
in Table \ref{fig:datasets-stats}.
Tables \ref{fig:sample-simple-questions}, 
\ref{fig:sample-medium-questions}, 
and \ref{fig:sample-complex-questions}
at the end of the paper
show three examples of each of these classes.

\begin{table}[ht]
\caption{Basic statistics of the sets of queries \cite{nascimento2024a}.}
\centering
\includegraphics[width=\linewidth]{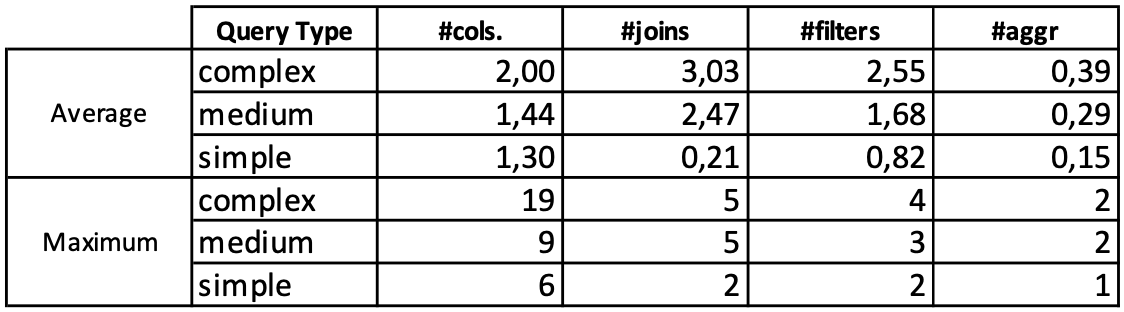}
\label{fig:datasets-stats}
\end{table}

\subsubsection{The Set of Partially Extended Views}
\label{sec:views}

The benchmark also includes
a set of \textit{partially extended views} \cite{nascimento2024b}
that rename table and column names of the relational schema 
to end users' terms.
Such views also have new columns that pre-define joins 
that follow foreign keys
and import selected columns from the referenced tables
to facilitate SQL query compilation.

They are maintained in the benchmark 
since they were used in the earlier experiments reported 
in Lines 1 to 3 of Table \ref{tab:results}.

\subsection{Evaluation Procedure}
\label{subsec:evaluation_metric}

The experiments used an automated procedure to compare
the \textit{predicted} and the \textit{ground-truth} SQL queries,
entirely based on column and table values, 
and not just column and table names.
Therefore, a text-to-SQL tool may generate SQL queries over the relational schema or any set of views,
and the resulting SQL queries may be 
compared with the ground-truth SQL queries
based on the results returned.
The results of the automated procedure
were manually checked to eliminate false positives and false negatives.
The reader is referred to \cite{nascimento2024a} for the details.

\subsection{Experiments with Schema Linking}
\label{sec:experiments-schema-linking}

\subsubsection{Experimental Setup}
\label{sec:setup-schema-linking}

The first set of experiments evaluated 
several alternatives for performing the schema linking task.

The experiments adopted the benchmark described in Section \ref{sec:benchmark}.
For each NL question,
the ground-truth minimum sets of tables necessary to answer the NL question
is the set of tables in the FROM clause of the ground-truth SQL query.
The experiments used GPT-3.5 Turbo and GPT-4, 
but only the results obtained with GPT-4 were noteworthy.

Table \ref{tab:schema_linking_res} 
presents the results of the experiments
for the following alternatives: 
\begin{enumerate}
\item \textit{(LLM)}: A strategy that prompts an LLM  
with $Q_N$ and $S$ to find the set of tables $S'$.
\item \textit{(DANKE)}: 
A strategy that, first, 
uses DANKE to extract a set of keywords $K$ from $Q_N$, 
and then extracts the set of tables $S'$ from 
the information associated with $K$ in DANKE's dictionary.
\item \textit{(LLM+DFE)}: A strategy that, first,
finds a set of examples $T$ from $D$ using $Q_N$,
and then prompts an LLM  
with $Q_N$, $S$, and $T$ to find the set of tables $S'$.
\item \textit{(LLM+DANKE)}: 
A strategy that, first,
uses an LLM to extract a set $K$ of keywords from $Q_N$,
and then extracts the set of tables $S'$ from 
the information associated with $K$ in DANKE's dictionary.
\item \textit{(LLM+DANKE+DFE)}: 
A strategy that, first,
uses an LLM to extract a set $K$ of keywords from $Q_N$,
retrieves the information associated with $K$ from DANKE's dictionary, creating a set $K_M$,
finds a set of examples $T$ from $D$ using $Q_N$,
and then prompts an LLM  
with $Q_N$, $S$, $K_M$, and $T$ to find the set of tables $S'$.
\item \textit{(Complete)}: The entire Schema Linking process.
\end{enumerate}

\subsubsection{Results}
\label{sec:results-schema-linking}

Table \ref{tab:schema_linking_res} presents the precision, recall, and F1-score for the experiments using the Schema Linking process. Briefly, the results show that:
\begin{enumerate}
\item \textit{(LLM)}: 
Alternative 1 obtained an F1-score of 0.851.
It had a performance poorer than Alternative 2, which used just DANKE.
\item \textit{(DANKE)}: 
Alternative 2 obtained an F1-score of 0.900.
Note that DANKE achieved a better result
than Alternatives 1 and 3 (which do not use DANKE),
although DANKE does no syntactic or semantic processing 
of the user question $Q_N$, 
and may incorrectly match terms in $Q_N$
to terms in the database schema or to data values.
\item \textit{(LLM+DFE)}: 
Alternative 3 obtained an F1-score of 0.868.
The use of DFE improved the results achieved by Alternative 1, 
but the results were still lower than those of Alternative 2.
\item \textit{(LLM+DANKE)}: 
Alternative 4 increased the F1-score to 0.930.
Enriching the prompt with the keywords extracted 
by DANKE from $Q_N$ 
yielded consistent improvements in both precision and recall.
This is due to DANKE's ability to find references to column values, 
associating them with the table/column where the value occurs. 
This feature allowed DANKE to find implicit references 
that were previously impossible for the LLM to discover
since it had no knowledge about the database instances.
\item \textit{(LLM+DANKE+DFE)}: 
Alternative 5 increased the F1-score to 0.950.
\item \textit{(Complete -- GPT-4)}: 
The complete Schema Linking process achieved an F1-Score of 0.996,
the best result. 
Using the LLM to extract keywords from $Q_N$ 
improved the results of DANKE. 
Although DANKE may still return incorrect terms, the LLM corrects them.
\item \textit{(Complete -- GPT-4o)}: 
Using GPT-4o resulted in a slight decrease in the F1-score to 0.995.
\end{enumerate}

\begin{table}[!ht]
    \centering
    \caption{Results for the schema linking alternatives (all with GPT-4, except the last line).}
    \label{tab:schema_linking_res}
    \scriptsize
    \begin{tabular}{|c|l|l|l|l|}
\hline
\# & \textbf{Method} &  \textbf{Precision} &   \textbf{Recall} &  \textbf{F1-score} \\ \hline
1 & LLM &   0.864 & 0.886 &  0.851 \\ \hline
2 & DANKE &   0.860 & 0.983 &  0.900 \\ \hline
3 & LLM+DFE &   0.940 & 0.843 &  0.868 \\ \hline
4 & LLM+DANKE &   0.930 & 0.930 &  0.930 \\ \hline
5 & LLM+DFE+DANKE &   0.993 & 0.983 &  0.950 \\ \hline 
6 & Complete -- GPT-4 &   0.993 & \textbf{1.000} &  \textbf{0.996} \\ \hline
7 & Complete -- GPT-4o & \textbf{1.000} & 0.995 & 0.995 \\ \hline
    \end{tabular}
\end{table}

In general, these results show that DANKE, together with the LLM,
performed effectively in the Schema Linking process
for NL questions. 
Considering that the complete Schema Linking process 
achieved a recall of 1.0, 
it returned all tables required to answer each NL question.
Thus, the Schema Linking process does not impact 
the SQL Query Compilation step, 
although the extra tables 
may create distractions for the LLM 
(see Section \ref{sec:experiments-compilation}).

\subsection{Experiments with SQL Query Compilation}
\label{sec:experiments-compilation}

\subsubsection{Experimental Setup}
\label{sec:setup-compilation}

The experiments were based on 
LangChain SQLQueryChain\footnote{\url{https://docs.langchain.com}}, 
which automatically extracts metadata from the database, 
creates a prompt with the metadata and passes it to the LLM. 
This chain greatly simplifies creating prompts 
to access databases through views
since it passes a view specification 
as if it were a table specification.

The experiments executed
the 100 questions introduced in Section \ref{sec:questions}
in nine alternatives:
\begin{enumerate}
\item \textit{(Relational Schema)}: 
SQLQueryChain executed over the relational schema 
of the benchmark database.
\item \textit{(Partially Extended Views)}: 
SQLQueryChain executed
over the partially extended views of the benchmark database.
\item \textit{(Partially Extended Views and DFE)}: 
SQLQueryChain executed over the partially extended views 
using only the DFE technique.
\item \textit{(Partially Extended Views, DFE, and Question Decomposition)}: 
SQLQueryChain executed
over the partially extended views,
using Question Decomposition and the DFE technique.
\item \textit{(The Proposed Text-to-SQL Strategy -- GPT-4)}: 
The proposed Text-to-SQL Strategy, using GPT-4-32K.
\item \textit{(The Proposed Text-to-SQL Strategy -- GPT-4o)}: 
The proposed Text-to-SQL Strategy, using GPT-4o.
\item \textit{(The Proposed Text-to-SQL Strategy -- LLaMA 3.1-405B-Instruct)}: 
The proposed Text-to-SQL Strategy, using LLaMA 3.1-405B-Instruct.
\item \textit{(The Proposed Text-to-SQL Strategy -- Mistral Large)}: 
The proposed Text-to-SQL Strategy, using Mistral Large.
\item \textit{(The Proposed Text-to-SQL Strategy -- Claude 3.5-Sonnet)}: 
The proposed Text-to-SQL Strategy, using Claude 3.5-Sonnet.
\end{enumerate}

Alternatives 1--6 ran on the OpenAI platform, 
and Alternatives 7--9 on the AWS Bedrock platform.

Also, recall that:
\begin{itemize}
\item GPT-4-32K has a context window of 32K tokens.
\item GPT-4o has a context window of 128K tokens.
\item Llama 3.1-405B Instruct has 405B parameters
and a context window of 128K tokens.
\item Mistral Large has 123B parameters
and a context window of 32K tokens. 
\item Claude 3.5 Sonnet has 175B parameters 
and a context window of 200K tokens.
\end{itemize}
\subsubsection{Results}
\label{sec:results-compilation}

\begin{table*}[!ht]
\centering
\caption{Summary of the results.}
\label{tab:results}
\includegraphics[width=1.0\linewidth]{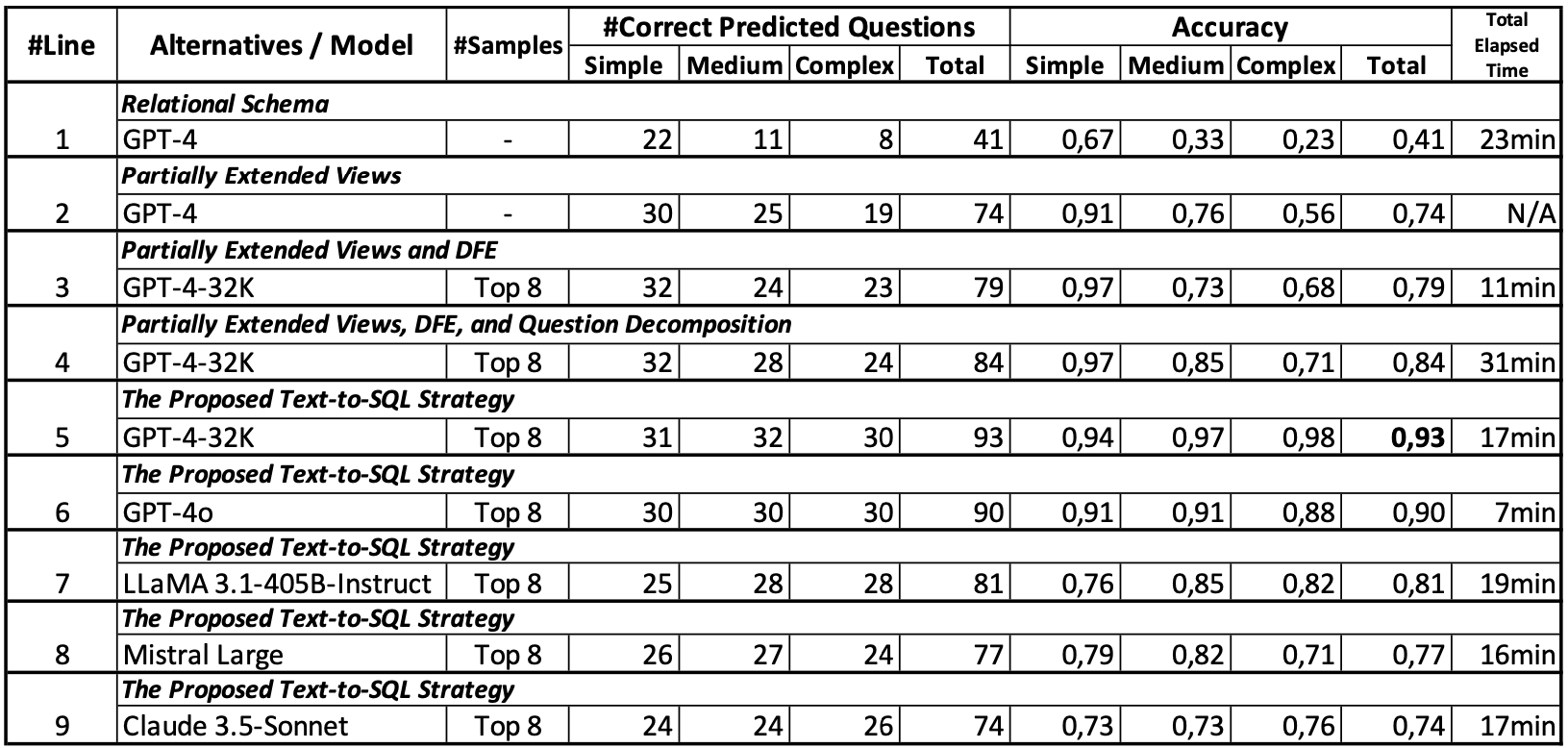}
\end{table*}

Table \ref{tab:results} summarizes the results
for the various alternatives.
Columns under ``\textbf{\#Correct Predicted Questions}''
show the number of NL questions per type
correctly translated to SQL
(recall that there are 33 simple, 33 medium, and 34 complex NL questions, with a total of 100); 
columns under ``\textbf{Accuracy}''
indicate the accuracy results per NL question type 
and the overall accuracy;
the last column shows the total elapsed time 
to run all 100 NL questions.

The results for Alternatives 1, 2, and 3
were reported in \cite{nascimento2024a,nascimento2024b,coelho2024},
respectively.
They are repeated in Table \ref{tab:results}
for comparison with the results of this paper.

The results for Alternative 4 
show that Question Decomposition
produced an improvement in total accuracy from 0.79 to 0.84.
This reflects the diversity of examples passed to the LLM when
they are retrieved for each sub-question, 
as already pointed out in \cite{Oliveira2024}.

The results for Alternative 5 
show that the key contribution of this paper,
the text-to-SQL strategy described in 
Section \ref{sec:prompt-strategy},
indeed leads to a significant improvement in the total accuracy for 
the case study database,
as well as the accuracies for the medium and complex NL questions.

The results for Alternative 6 
indicate a slight decrease in the total accuracy to 0.90
when GPT-4o is adopted,
possibly due to the non-deterministic behavior of the models.
However, while GPT-4-32K took 17 minutes to run all 100 questions,
GPT-4o took only 7 minutes.

The results for Alternatives 7--9 show a decrease in the total accuracy
to 0.81\%, 0.77\%, and 0.74\%, respectively,
with a much higher total elapsed time when compared with GPT-4o,
but comparable to that of GPT-4-32K.

\subsubsection{Discussion}
\label{sec:discussion}

In all alternatives that did not resort to DANKE,
the LLM had all the Schema Linking burden,
and had to synthesize all the joins 
required to process the NL question correctly.
By contrast, the strategy
of Section \ref{sec:prompt-strategy},
by using DANKE,
alleviated these burdens.

DANKE also helped the LLM with ambiguous questions 
both during the Schema Linking 
and the SQL Query Compilation processes.

In a few cases, Schema Linking returned
more tables than required.
But, in most of these cases,
the SQL Query Compilation process
was not jeopardized;
in only one case, the extra tables and consequently
the extra columns in the view 
led the LLM to confuse the choice of columns,
and synthesize an incorrect WHERE clause.

The results in Table \ref{tab:results} show that 
the proposed strategy (Line 5) 
correctly processed four more medium NL questions 
and six more NL complex questions than 
the previous best strategy (Line 4).
However, these results hide the fact that
the proposed strategy processed four complex NL questions
that none of the strategies 
previously tested on the same database and set of questions
have correctly handled,
including C3 and DIN.
Table \ref{fig:sample-complex-questions}
shows three such NL questions.
For example, in Question 29, 
the previous strategies 
did not correctly synthesize the required join,
whereas the view created in Step 1 of the SQL Query compilation process (see Section \ref{sec:query-compilation})
indeed includes such join,
making it easier for the LLM to compile the required SQL query.
In Question 93, all previous strategies failed
to remap the installation name ``E176'' in the user question 
to the installation name ``E-176'' stored in the database;
the proposed strategy used DANKE's matches to correct this problem.

Similar observations 
are valid for the medium NL questions.
For example, the previous strategies 
did not correctly process Question 24 
in Table \ref{fig:sample-medium-questions},
whereas the proposed strategy did,
using DANKE's matches.

An analysis of the NL questions 
compiled into incorrect SQL queries
uncovered that the proposed strategy  
failed for two basic reasons: 
(1) the semantics of a term of the NL question
was mapped into an incorrect SQL filter
(i.e., due to a \textit{semantic mismatch}); 
and (2) a term of the NL question 
was associated with an incorrect column name.

As for the other models -- 
Llama 3.1-405B Instruct, Mistral Large, and Claude 3.5 Sonnet --
the most common source of error was the use of the CONTAINS function, 
which requires the target column to be indexed, 
but this was not always the case;
the correct filter would have to use LIKE.

\section{\uppercase{Conclusions}}
\label{sec:conclusions}

This paper proposed a strategy
to compile NL questions into SQL queries,
especially questions that require complex filters and joins,
that leverages the services provided by DANKE,
a database keyword search platform.

The paper detailed how the schema-linking process
can be improved with the help of the keyword extraction service
that DANKE provides.
Then, it showed how DANKE can be used 
to synthesize a view that captures the joins required
to process an input NL question
and thereby simplify the SQL query compilation step.

Both the schema-linking and 
the SQL query compilation processes
use a dynamic few-shot technique,
based on a synthetic dataset constructed from the database.
Section \ref{sec:synthetic-dataset} described
a technique for constructing the synthetic dataset
that improves the technique introduced in \cite{coelho2024}.

The paper included experiments 
with a real-world relational database 
to assess the performance of the proposed strategy.
The results in Section \ref{sec:experiments-schema-linking}
showed that the precision and recall of the schema-linking process
indeed improved with the help of the keyword extraction service
that DANKE provides.
The discussion in Section \ref{sec:experiments-compilation}
suggested that creating a view with the help of DANKE 
also helped with the SQL query compilation process.
In conjunction, these results indicated that
the proposed strategy achieved a total accuracy 
in excess of 90\% over 
a benchmark built upon a relational database
with a challenging schema
and a set of 100 questions carefully defined
to reflect the questions users submit
and to cover a wide range of SQL constructs.
The total accuracy was much higher
than that achieved by 
SQLCoder, LangChain SQLQueryChain, C3, and DIN+SQL
on the same benchmark,
as reported in \cite{nascimento2024a}.

As future work, the proposed strategy
should be tested and compared against other strategies 
using additional databases and test questions.
This effort depends, however, on working with 
real-world databases that are available, populated, 
and with good documentation.
As a first step, the strategy has already been applied 
to other databases that are in production and 
to the openly available Mondial database,
with positive results.

A second demand is to address the problem that 
Natural Language questions are intrinsically ambiguous.
The use of DANKE's matching process helps, 
but it should be complemented with a different approach,
perhaps incorporating the user in a disambiguation loop.

Finally, it should also be stressed
that the proposed strategy 
and the synthetic dataset construction process are generic,
but costly to set up.
They should be considered when 
it is worth investing in an NL interface 
for a serious database where accuracy is at stake.

\section*{\uppercase{Acknowledgements}}
This work was partly funded 
by FAPERJ under grant E-26/204.322/2024; 
by CNPq under grant 302303/2017-0; 
and by Petrobras, under research agreement 2022/00032-9 
between CENPES and PUC-Rio.

\bibliographystyle{apalike}
{\small
\bibliography{references}}

\begin{table*}[ht]
    \centering
        \caption{Sample NL simple questions and their SQL golden standard (anonymized).}
    \label{fig:sample-simple-questions}
    \includegraphics[width=\linewidth]{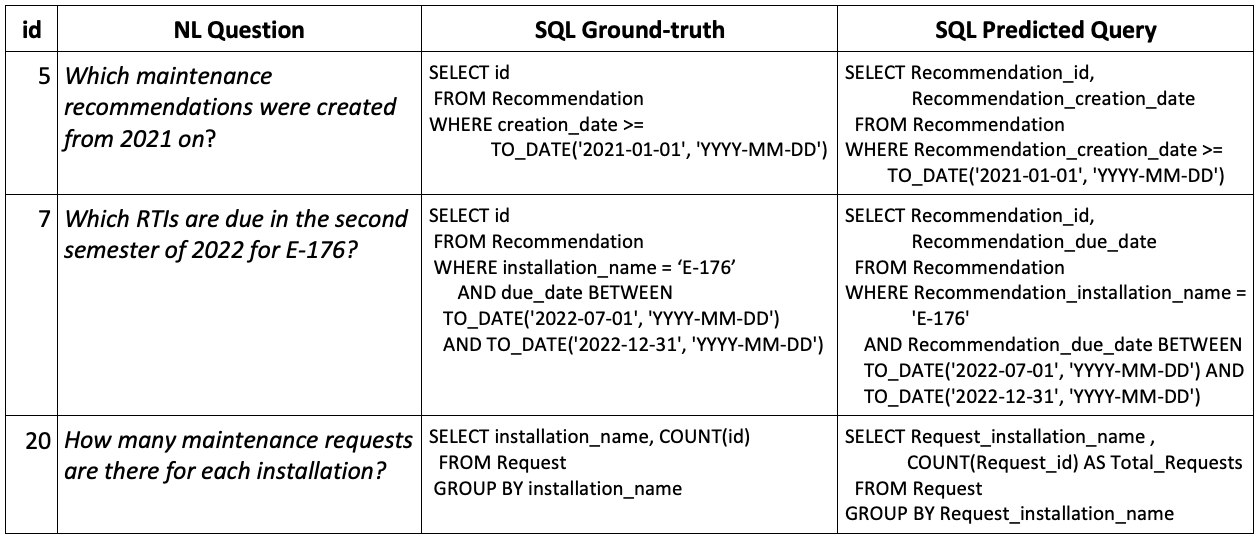}
\end{table*}
\begin{table*}[ht]
    \centering
    \caption{Sample NL medium questions and their SQL golden standard (anonymized).}
    \label{fig:sample-medium-questions}
    \includegraphics[width=\linewidth]{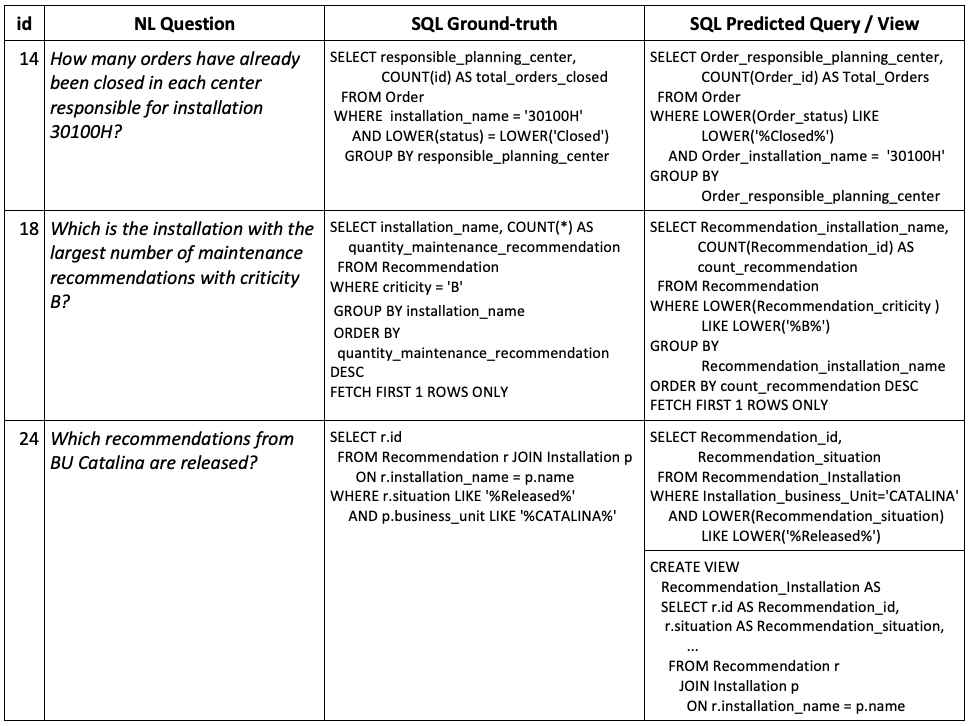}
\end{table*}
\begin{table*}[ht]
    \centering
    \caption{Sample complex NL questions and their SQL golden standard (anonymized).}
    \label{fig:sample-complex-questions}
    \includegraphics[width=\linewidth]{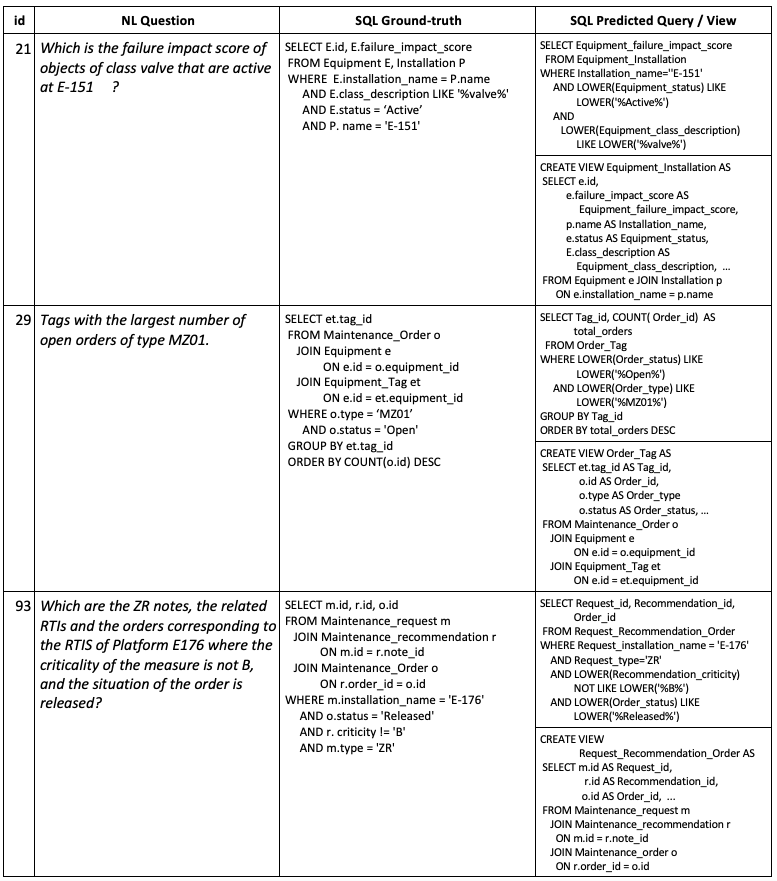}
\end{table*}

\end{document}